\documentstyle[aps,multicol,pre]{revtex}
 \title{Surface rearrangement at complex adsorbate-substrate 
 interfaces.
}

 \author{E.V. Vakarin  and J.P. Badiali
}
 \address{LECA
 ENSCP-UPMC, 11 rue P. et M. Curie, 75231 Cedex 05, Paris, France\\
 }

\begin{document}
 \maketitle
 \begin{abstract}
On the basis of the information theory approach we propose a novel 
statistical scheme for analyzing the evolution of coupled 
adsorbate-substrate systems, in which the substrate undergoes
the adsorbate-induced transformations. 
A relation between the substrate
morphology and the adsorbate thermodynamic state is established. 
This allows one to estimate the surface structure in terms of incomplete
experimental information and the one concerning the adsorbate 
thermodynamic response to the structural modifications.  
\end{abstract}

\begin{multicols}{2}

\section{Introduction}
Adsorbate-induced changes in the substrate morphology is a quite common
phenomenon in nature. This concerns with in-plane and out-of-plane 
surface modifications, such as reconstruction or roughening (see
\cite{Persson} for a recent review).

For simple crystalline surfaces (single crystals) there are well-developed 
Hamiltonian models dealing theoretically with the adsorbate-induced 
restructuring effects\cite{PRL}. Focusing on the roughness, the 
solid-on-solid models\cite{ZhR,PRBr} are appropriate for the description of 
the surface morphology under the influence of adsorption. Nevertheless, even
in this "traditional" domain there are certain unsolved problems
with the surface structure determination\cite{woodruff}. Therefore, one 
has to work under conditions of incomplete information.  

Moreover, many  
adsorbents do not fall into this class because of their intrinsic 
complexity. The latter appears as a consequence of the surface disorder, 
heterogeneity\cite{zgrablich} or polydispersity. 
Glassy\cite{H-glass}, 
granular, amorphous\cite{Julien}, or porous materials, like 
aerogels\cite{aerogel1,aerogel},  could serve as examples.
The morphology of such materials is conventionally specified by
a probability distribution (e.g. site energies, pore or grain sizes). In 
many cases the distribution itself is unknown. For randomly heterogeneous 
surfaces the information comes from scattering experiments or STM images.
In the case of three-dimensional structures one has
to rely upon results of indirect measurements (such as nitrogen adsorption,
mercury porosimetry, thermoporometry, etc). These techniques give a quite
limited information from which the probability distribution should be
determined. This problem is beyond the scope of the standard
statistical mechanics, which starts from a microscopic model. 
  
Actually, the situation is complicated by the fact that the
surface morphology\cite{galle} and/or energetics\cite{nieto} change in the 
course of adsorption. This is quite common phenomenon, occurring, for 
instance, during the aerogel characterization\cite{aerogel1,aerogel} or in
the course of various insertion processes\cite{PRB,JPCB}. In this 
respect the problem is essentially as follows.
One has to describe the surface evolution from a poorly characterized
initial (e. g. clean surface)  to a final (adsorbate-induced) state.
In addition, the adsorbate thermodynamics, coupled to the surface
morphology, should be calculated.  The inverse problem, that is, restoring
the clean surface characteristics from its morphology in the covered state,
is also important.

In this paper we make an attempt to deal with such problems combining
the information theory approach\cite{jaynes,haken} 
and the standard statistical thermodynamics. The adsorbate-substrate
system can be split into subsystems with different levels of
description. 
One is a dynamic subsystem (the adsorbate), which evolves according to a 
Hamiltonian, containing unknown parameters (local roughness of the surface).
The surface is considered as a stochastic system. Its morphology (e.g. 
step configuration) is specified by a probability 
distribution. This combined scheme looks quite similar to the so-called 
superstatistics\cite{beck}, that describes the statistics in systems with 
fluctuating intensive quantities. The key difference is that in our case the 
probability distribution is unknown. It has to be inferred from a limited
number of constraints. Recently such a maximum-entropy procedure has been
applied to a description of velocity fluctuations in turbulent 
fluids\cite{sattin} and the strain fluctuations in heterogeneous 
intercalation systems\cite{EACTA}. 
 In our case the adsorbate can influence the surface roughness  and 
 their 
coupling determines both the thermodynamic behavior and the shape of the 
probability distribution. It is 
essential that   one necessarily deals 
with two entropic impacts\cite{CEJP}: the thermodynamic entropy (due to the 
adsorbate) and the information entropy (due to the 
probability distribution). Constructing from these two terms a suitable 
entropy measure, we investigate the evolution of the surface morphology and 
the adsorbate thermodynamics. The main purpose of this paper is to find a 
relation between the surface structure and the adsorbate thermodynamic
response to these structural modifications. This would allow one to
gain an information on the surface following the adsorbate behavior.
\section{Clean surface}
In this section we demonstrate how the maximum entropy formalism
can be applied to inferring a distribution based on a limited information.
For concreteness we consider a solid-like semi-infinite substrate (see 
Figure~1). The surface (substrate) morphology is represented by a collection 
of steps $\{h_i\}$ growing on discrete sites $i=1..N_s$, with 
$-\infty<h_i<\infty$, describing a deviation from the flat geometry.
A sketch of the surface profile is given in Figure~1. 

It is assumed 
that, because of its complex nature, the surface cannot be described 
deterministically, i.e. in terms of a Hamiltonian governing the substrate 
evolution. Thus we have to accept a probabilistic description, 
focusing on a distribution $P(\{h_i\})$ of the step configurations. However, 
the distribution is unknown. The only available information concerns with a 
set of experimental tests determining\cite{note1}, for instance, the
average roughness   
\begin{equation}
\label{constraint2} 
H_0=\int (dh_i) P(\{h_i\})\sum_i h_i^2
\end{equation}

For simplicity the steps are taken as spatially non-correlated 
$P(\{h_i\})=\prod_i P(h_i)$. Then $\sum_i h_i^2=N_sh^2$, where $N_s$ is the 
number of steps. Therefore, $P(h_k)$ is a probability to find a step of 
magnitude $h$ at a randomly chosen surface site $k$.    

Having such a restricted information on the substrate, one can 
infer its statistics within the information theory approach
\cite{jaynes,haken}.
The inference procedure is based on a given entropy measure $S_I$, which
is taken here in the Shannon form
\begin{equation}
S_I=-\sum_k\int dh_k P(h_k) \ln P(h_k)
\end{equation}
This entropy is a measure of our uncertainty on the system, in the sense
that $S_I$ increases with broadening $P(h_k)$. 
The least biased normalized distribution, compatible with the constraint 
(\ref{constraint2}) is obtained by extremizing the Lagrangian
\begin{eqnarray}
\Lambda=S_I -
\nu \sum_k\left[ \int dh_k P(h_k)-1\right] -\\\nonumber
\lambda_0\left[\sum_k\int dh_k P(h_k)h_k^2-H_0 \right] 
\end{eqnarray}
This leads to a gaussian distribution
\begin{equation}
P_0(h_i)=\frac{e^{-\lambda_0h_i^2}}
{\int dh_i e^{-\lambda_0h_i^2}}
\end{equation}
The Lagrange multiplier $\lambda_0$ can be determined from the constraint
(\ref{constraint2}) $\lambda_0=1/(2H_0)$. We do not discuss here how precise
is this description. Having more information, e.g. 
$\langle h^4 \rangle$, we can follow the same scheme in order to obtain a 
refined distribution. 
\section{Surface-adsorbate coupling}
Being exposed to a fluid, the surface can adsorb the fluid particles
on the sites (see Fig.~1). Having information\cite{note1} on the roughness 
$H$ in the presence of adsorption, one can infer the step distribution 
maximizing $S_I$. Obviously, the distribution is also gaussian 
\begin{equation} 
P(h_i)=\frac{e^{-\lambda h_i^2}}
{\int dh_i e^{-\lambda h_i^2}}
\end{equation}
but with a different width $\lambda=1/(2H)$. Therefore, if $H$ is
close to $H_0$, then one would conclude that $P(h_i)=P_0(h_i)$ and 
the adsorbate-induced effects
are negligible. This can be the case if the surface is quenched. As is 
discussed above, the surface may change its morphology even if the 
average roughness remains unchanged. In other words, we do not have new
{\it a priori} information if $H=H_0$. Then, starting with the same entropy 
measure $S_I$, we arrive at a trivial (but misleading) result:  
$P(h_i)=P_0(h_i)$. This defect can be avoided in (at least) two ways.
One might need to compare higher moments (for instance, the 
average of $h_i^4$). This would require additional tests\cite{note1}. 
Alternatively, we may refine the entropy measure, taking into account 
the adsorbate-induced effects. This requires to make a link between the 
surface structure and the adsorbate properties.

The adsorbate is considered as a dynamic
subsystem, whose coupling to the surface is given by a Hamiltonian
${\cal{H}}(\{t_i\}|\{h_k\})$. We assume that the 
surface evolves in time much slower than the adsorbate, that reaches the 
equilibrium while the surface is practically unchanged.  Tracing over the 
set of occupation numbers $\{t_i\}$ at a given step configuration we can 
calculate the partition function
$$
Z(\tau|\{h_k\})=\sum_{\{t_i\}} e^{-\beta {\cal{H}}(\{t_i\}|\{h_k\})}
$$
where $\beta=1/k_B T$ is the inverse temperature, and 
$\tau$  is a thermodynamic variable (e. g., chemical potential or coverage, 
see below). Then the adsorbate thermodynamics, for instance, the 
conditional entropy $S_T(\tau|\{h_k\})$, can also be calculated. 

In what follows this quantity will play an essential role. Therefore, 
it is important to clarify its meaning. As a function of $\tau$ the entropy 
is the conventional thermodynamic state function. It 
 is a measure of the adsorbate disorder for a given surface structure
 $\{h_k\}$. On the other hand, if $\tau$ is fixed (e. g. fixed adsorbate 
 pressure), then $S_T$ as a function of $\{h_k\}$ should be considered
 as an effective potential, selecting a preferable step configuration --
 the one which maximizes $S_T$. 
  
Since we are dealing with adsorbate-induced effects, then 
$S_T(\tau|\{h_k\})$
should be relevant to the surface structure evolution. Therefore  
we refine the entropy measure  
\begin{equation}
\label{Sigma}
\Sigma = S_I+ \kappa\int (dh_k) P(\{h_k\})S_T(\tau|\{h_k\})
\end{equation}
which takes into account the fact that some part of our uncertainty on the 
surface state is due to the adsorbate thermodynamic state 
$S_T(\tau|\{h_k\})$. In what follows we assume that the steps do not 
correlate through the adsorbate 
$S_T(\tau|\{h_k\})=\sum_k S_T(\tau|h_k)$. Imlicitly this means a restriction 
on the Hamiltonian ${\cal{H}}$ (see below).
 
 The parameter $\kappa$ is a measure of the adsorbate 
influence on the step distribution. In other words, $\kappa$ is a degree
of the surface quenching, such that $\Sigma \to S_I$ as $\kappa \to 0$
and we return to the clean surface problem. Determination of $\kappa$
is a separate problem that might require additional tests. For instance,
comparing the surface step creation energy $\varepsilon$ and the adsorbate
binding energy excess $\gamma$, we may estimate 
$\kappa=\gamma/(\varepsilon+\gamma)$. However, in this study $\kappa$
is taken as a parameter.

Therefore, we consider the Lagrangian 
\begin{equation}
\Lambda'=\Lambda+\kappa\sum_k\int dh_k P(h_k)S_T(\tau|h_k)
\end{equation}
that must be extremized. This 
procedure leads to 
\begin{equation}
\label{Pnew}
P_{\kappa}(h_i)=\frac{e^{-\lambda_{\kappa} h_i^2+\kappa S_T(\tau|h_i)}}
{\int dh_i e^{-\lambda_{\kappa} h_i^2+\kappa S_T(\tau|h_i)}}
\end{equation}
The multiplier $\lambda_{\kappa}$ should be determined from
the constraint on the roughness $H$. The term $\lambda_{\kappa} h_i^2$
partially keeps traces of the surface preparation (a given roughness $H$).
The second term $\kappa S_T(\tau|h_i)$ corresponds to the adsorbate-induced
effects. If $S_T$ is an analytic function, 
then  we may expand\cite{note2} 
$$
S_T(\tau|h_i)=\sum_n A_n(\tau)h_i^n
$$ 
Therefore, even without resorting to any specific
model, it is clear that $P_{\kappa}(h_i)$ is potentially more informative 
than $P(h_i)$ as the former contains higher order terms (in $h_i$). As is 
noted above, in order to obtain a similar result staring from $S_I$, one 
needs additional information. Therefore, refining the entropy measure, 
we restore (at least partially) this information. In particular, we can
calculate a global change of the surface structure in comparison to
the clean surface case. This can be done using the symmetric 
Kullback-Leibler information measure 
\begin{equation}
K=\sum_i\int dh_i \left[ P_{\kappa}(h_i)-P_0(h_i) \right]
\ln \left[
\frac{P_{\kappa}(h_i)}{P_0(h_i)}
\right]
\end{equation}
which tells us how much these two distributions are different.
$K$ is found to be 
\begin{eqnarray}
K=-(\lambda_{\kappa}-\lambda_0)
(H-H_0)\\\nonumber
+\kappa \sum_i
\left[
\langle S_T(\tau|h_i) \rangle_{\kappa}-\langle S_T(\tau|h_i) \rangle_0
\right]
\end{eqnarray}
where the angular brackets 
$\langle ... \rangle_0$ and $\langle ...\rangle_{\kappa}$
denote the averages taken with the distributions
$P_0(h_i)$ and $P_{\kappa}(h_i)$, respectively. Therefore, even if 
$H=H_0$, the global variation of the surface structure, induced by adsorption
is proportional to the change of the adsorbate thermodynamic state.
The state $\langle ... \rangle_0$ does not
necessarily correspond to the clean surface, it could be any suitable
reference state. In the same way one can estimate the information
gain due to the refinement of the distribution (replacement $P(h_i)$
by $P_{\kappa}(h_i)$). 
This result is quite general, it does not depend on the microscopic details
underlying $S_T(\tau|h_i)$. Nevertheless, in order to study the distribution
itself we need to consider a microscopic model.

\section{Illustration}
In order to demonstrate what kind of local changes are induced by adsorption,
as an illustration we choose one of the simplest adsorption models -- 
the non-interacting lattice gas with a height-dependent binding energy 
\cite{SUSC,CPLR}) 
\begin{equation}
{\cal{H}}(\{t_i\}|\{h_k\})=-\sum_i (\mu +\epsilon(h_i))t_i
\end{equation} 
We are aware that the model itself is quite "rough" in application 
to real adsorbates. It neglects the lateral interactions and, therefore,
is inadequate for the description of the adsorbate critical behavior 
(such as liquid-gas or order-disorder transitions). Nevertheless above
the critical temperature this model is qualitatively correct. In addition
it is exactly solvable such that one can be sure that our implementations
are not artifacts of approximations (e.g. mean field). Moreover, we do
not take into account that the roughness increases the surface area, 
allowing for more than a monolayer coverage\cite{CPLR}. Thus $t_i=0$ or $1$ 
independently of $h_i$. 

The entropy of this simplified model is given by (the role of $\tau$ is 
played by the chemical potential $\mu$) 
\begin{equation}
\label{ST}
S_T(\mu|h_i)=-\theta(h_i)\ln[\theta(h_i)]-[1-\theta(h_i)]\ln[1-\theta(h_i)]
\end{equation}
 Here the coverage is Langmuirian
\begin{equation}
\label{thetah}
\theta(h_i)=\frac{L e^{\beta \epsilon(h_i)}}
{1+L e^{\beta \epsilon(h_i)}}
\end{equation}
The adsorbate activity $L$ is related to the chemical potential
$\mu$ via $L=\exp(\beta \mu)$, and $\epsilon(h_i)$ is the height-dependent 
adsorbate binding energy.
\begin{equation}
\epsilon(h)=\epsilon_0+\gamma h_i^2
\end{equation}
A step can be viewed as a loosely coordinated site with the binding
energy larger\cite{SUSC,CPLR} than that for the flat surface.
Here $\gamma$ is the binding energy excess due to the local deviation
from the flat geometry. The binding to the flat surface $\epsilon_0$ is an 
irrelevant constant that can be absorbed into the definition of $L$.

In Figure~2 $S_T(\mu|h_i)$ is analyzed as a function of $h_i$ and
dimensionless activity $m=\beta(\mu+\epsilon_0)$. The entropy (as a
function of $m$) behaves in the usual way, exhibiting a single maximum
at some $m^*$ which depends on the magnitude of $h_i$ and $\gamma$. At 
relatively high adsorbate activities ($m>0$)  $S_T(\mu|h_i)$ is also 
single-peaked as a function of $h_i$, with the maximum located at $h_i=0$. 
This suggests a tendency towards random deviations from $h_i=0$.
This maximum transforms into a minimum when the activity changes its sign
($m<0$). Then the entropy develops two symmetric peaks at some
$|h_i^*|$ that depends on the magnitude of $m$ and $\gamma$. This means that
the position $h_i=0$ becomes unfavorable and the surface would tend 
to exhibit a random zig-zagging with the steps distributed near $\pm 
h_i^*$.  Therefore, considering $S_T(\mu|h_i)$ as an effective potential
for the step variables $h_i$, one expects qualitative changes in the
surface morphology with changing adsorbate activity $m$.  

In 
order to analyze (qualitatively) what this implies for the distribution 
$P_{\kappa}(h_i)$, we take $\lambda_{\kappa}$ as a parameter and set the 
normalization constant (the denominator in (\ref{Pnew})) equal $1$. Then we 
cannot calculate averages, but these simplifications do not change the shape 
of the distribution. Such a
reduced $P_{\kappa}(h)$ is plotted schematically in Figure~3. It is
seen that with increasing $\kappa$ the distribution changes from a 
gaussian ($\kappa=0$) to a tri-modal. This results from an interplay
between the adsorbate-induced effects and the tendency to keep a given
roughness.     
The curves coincide in the
large-$h$ region because the distribution is not normalized. For
quantitative purposes this defect can be removed by numerical integration
in eq.~(\ref{Pnew}). It is clear (see Fig.~1) that the adsorbed species
occupy the space of the corrugated surface layer\cite{SUSC}.   
Therefore the adsorbate is confined to
a non-planar slit, whose average thickness $T$ is determined by the surface
roughness $H$ (see Figure~1). Moreover, if $h_i$ gives a length scale, then
the surface roughness $H$ (the average of $\sum_ih_i^2$) is at the same time 
the surface area excess in comparison to a flat surface. Thus 
$P_{\kappa}(h_i)$ can be considered as a surface area distribution. Within 
this analogy we make a qualitative comparison to Monte Carlo 
results\cite{galle} for the surface area fraction in pores with rough walls 
(see the insets in Figure.~2). The right inset exhibits a peaked (nearly 
gaussian) distribution that corresponds to an empty pore. The left inset 
depicts the adsorbate-induced surface structure. It is seen that our 
approach describes quite well this physics (recall that $\kappa$ is a measure
of adsorbate-induced effects). We did not try to fit, because the 
models are not strictly identical. It is remarkable that our scheme
is capable of inferring the surface structure even without entering
the microscopic details concerning the surface. Moreover the simulation
results\cite{galle} are not much sensitive to the type of adsorbate-wall 
interaction (hydrophilic or hydrophobic, see Fig.~5 in \cite{galle}).
This suggests that the adsorbate-induced surface rearrangement is mainly
an entropic effect. This is coherent with our implications.

\section{Conclusion}
The evolution of coupled adsorbate-substrate system is investigated
in terms of a new approach combining the information theory and the
standard statistical mechanics. The approach is suitable for complex
systems, where, for a number of reasons, not all the microscopic details 
(e. g. a Hamiltonian) are available.

The approach establishes a clear correlation between the substrate
morphology (e. g. roughness) and the adsorbate thermodynamic state.
This implies a statistical scheme for surface characterization in the course 
of the adsorbate-substrate coupling, and  allows one to estimate the 
surface morphology based on the information concerning the adsorbate 
thermodynamic response. One can also predict the clean surface properties
based on its morphology under adsorption and vice versa. 
Thus, our scheme is, in some sense, complimentary to the recent (mainly 
simulational) surface characterization techniques 
\cite{zgrablich,galle,nieto}. It can also be applied to analyzing the 
experimental data on porous adsorbents \cite{aerogel1,aerogel}.

We would like to underline that the simple model
considered here serves just as an illustration. To be more realistic
in further applications we have to consider the adsorbate-adsorbate 
interactions and a correlation between the steps. Also, the experimentally 
available information ($H_0$ or $H$) is supposed to be quite crude. In
many cases of real surfaces  a more detailed 
statistics can be extracted (for instance, from STM images),
for example,- pair correlation $H_{ij}=\langle (h_i-h_j)^2\rangle$. 
Nevertheless, for three-dimensional matrices (like aerogels) in the 
majority of cases only indirect measurements are available. 
All these features can be easily incorporated into our approach.

Apart from the practical applications, our approach seems promising
in analyzing a class of coupled dynamic-stochastic systems, where the
stochastic subsystem is not perfectly quenched. Particularly interesting
is to study the adsorbate critical behavior under conditions when a porous
matrix responds to the adsorbate thermodynamics (e.g. helium absorption in
aerogels). 
 
\end{multicols}
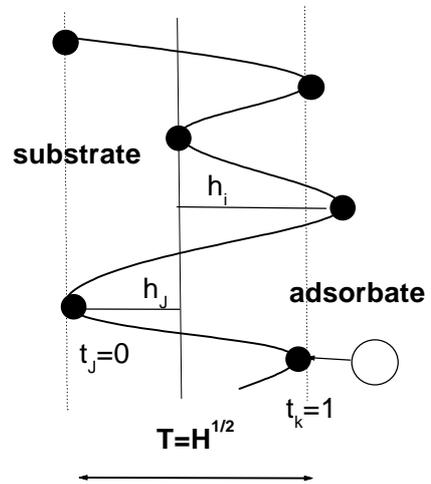
\begin{figure}
\caption{A sketch of the adsorbate-substrate interface. The black
dots mark the adsorption sites. The surface steps $h_i$ are defined
for each site. The solid line is a reference  from which all $h_i$ are 
counted.}
\end{figure}
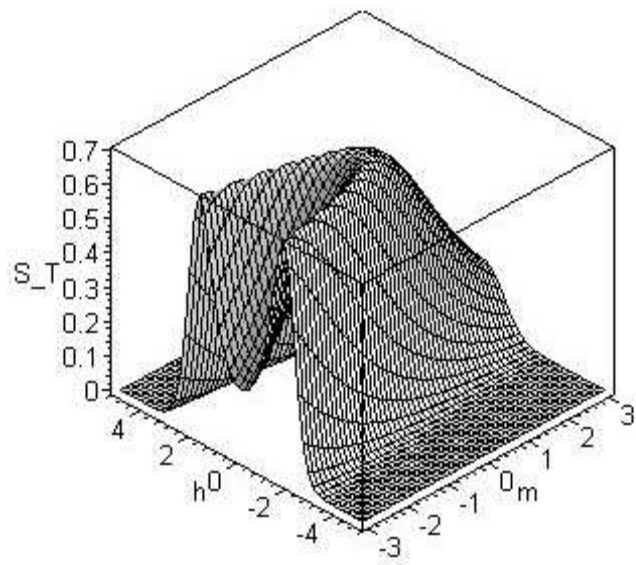
\begin{figure}
\caption{ The conditional entropy  $S_T(\mu|h_i)$ as a function of 
$h_i$ and dimensionless activity $m=\beta(\mu+\epsilon_0)$,   
$\beta \gamma=1$}. 
\end{figure}
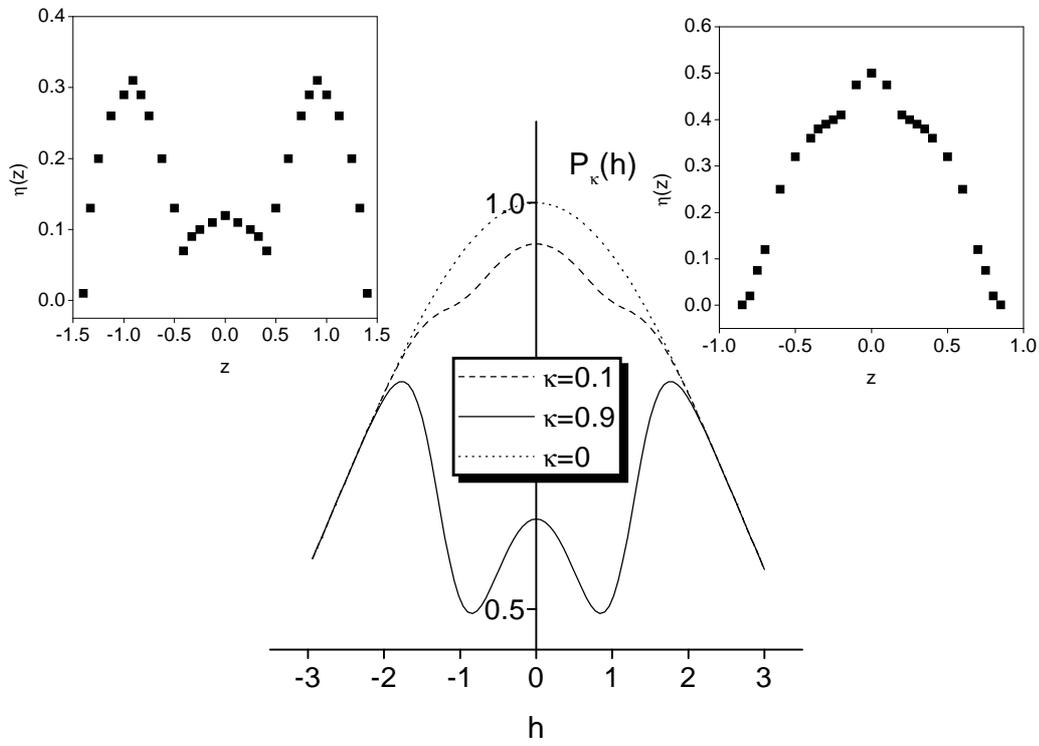
\begin{figure}
\caption{A non-gaussian step probability distribution, resulting
from eq.~(8) ($\lambda_{\kappa}=0.1$, $L=0.1$, $\beta\gamma=2$). The 
insets demonstrate the Monte Carlo results [11] for the surface area 
fraction in pores with rough walls. } 
\end{figure}

\begin{thebibliography}{99}
\bibitem{note1} Note, however, that we do not implement here any specific
 experimental method.
\bibitem{note2} In general, for interacting adsorbates, the entropy 
 expansion should contain terms like $\alpha_{ij}(h_i-h_j)^n$. Then the 
 steps would correlate through the adsorbate, even thought they were 
 uncorelated for the clean surface case.
\end{thebibliography}
\end{document}